# LOCK-IN TO COMMENSURATE STATES INDUCED BY A PERIODIC ARRAY OF NANOSCALE ANTI-DOTS IN Nb SUPERCONDUCTOR

A.A.Zhukov and P.A.J.de Groot

Department of Physics and Astronomy, University of Southampton, Southampton, SO17 1BJ, UK

V.V.Metlushko

Department of Electrical and Computer Engineering, University of Illinois at Chicago Chicago, IL 60607-0024

B. Ilic

School of Applied and Engineering Physics, Cornell University, Ithaca, NY 14853

We study the interactions of the vortex lattice with a periodic square array of holes in a superconducting Nb film. Using high resolution magnetic field measurements of electrical losses, extremely narrow states with a magnetic field width reaching 1% of the matching field value are found at the commensurate points. They are accompanied by pronounced harmonic generation in response to ac modulation of the magnetic field or current. We relate these sharp anomalies to a locked commensurate state of the Mott insulator type.

The interaction between a periodic elastic medium and a periodic pinning potential represents the behavior of many important physical systems. Besides the case of the vortex lattice interacting with a periodic artificial pinning array in type II superconductors, it includes the behavior of vortices in layered superconductors with intrinsic pinning[1], charge[2] and spin[3] density waves, mass density waves in superionic conductors[4], polarization density waves in incommensurate ferroelectrics[1], absorbed atomic layers on periodical substrate[5], magnetic bubble arrays[6] and the magnetically induced Wigner crystal in a two dimensional electron gas[7]. The easy control of vortex separation in superconductors by simply varying the external field and the large diversity in possible interaction manifolds and in intrinsic parameters makes the vortex system one of the most versatile experimental testing fields.

Modern technologies allow us to prepare samples with an efficient periodic system of pinning centres. Among them much interest is generated by arrays of holes with submicron length scales. Interaction of the vortex system with such a regular array of anti-dots results in a wide range of commensurability effects observed for integer or rational number of flux quanta per a pinning site[8-11]. Many interesting effects have been found, including the existence of multiquanta vortices on holes[9, 10], vortices taking interstitial positions[9, 12], Shapiro steps[13], field polarity dependent flux pinning[14], etc. Individual and multi-vortex pinning was considered in numerical simulations and resulted in a series of novel phases including multi-vortex and composite superlatice states such as the aligned dimer and trimer configurations at individual pinning sites[15, 16]. However, several aspects related to the commensurate-incommensurate transition and, even more important, the region of the locked commensurate state remains to be clarified. This state was shown to correspond to a Mott insulator state in the spectrum of quantum bosons [17, 18]. In our Letter we report the existence of extremely narrow states at matching fields. We relate these sharp anomalies to a locked commensurate state of the Mott insulator type.

An anti-dot structure was prepared in a Nb superconducting film with a thickness of 100nm using laser interferometric lithography[19, 20]. This method allows us to approach an ideal periodicity on the scale of whole sample. The high quality of the arrays prepared in this manner was evident from the large number of commensurate states observed[20, 21]. In the sample used in our investigations a periodic square lattice of holes penetrating full film thickness with a diameter of 400 nm and a period of 1100 nm was prepared. The onset of the superconducting transition was found to be $T_c$ = 7.88 K, and the resistive transition width between the levels of 10-90% is 0.09 K. The smaller value of $T_c$ in comparison with bulk Nb is typical[21, 22] and probably originates from a background oxygen contamination of the thin film. Plain film demonstrated similar superconducting transition. The resistance was measured by a standard four probe technique using a square modulated current with a frequency of 68 Hz. The samples have an elongated rectangular shape and a width of 1-2 mm. The distance between voltage probes was 1-2 mm. To reveal the locked state we employed a modulation technique. We apply either a sinusoidal magnetic field of 1 mOe - 1 Oe amplitude to the sample with a dc current in the range 1 $\mu$A – 1 mA or use a sinusoidal modulation of the current with a frequency in the range 10 Hz-10 kHz. Harmonic analysis is used as a sensitive indicator of phase transitions, which is usually accompanied by strong non-linearities. The measurements were made near the superconducting transition and hence our temperature stability below 0.5 mK was crucial for this experiment.

Fig.1a shows a typical behavior of resistance vs. magnetic field R(H). Characteristic sharp dips are observed at matching fields[22, 23]. The large regularity of the hole array structure resulted in a large number of matching anomalies. At low temperatures we detect features up to n=11. From the matching fields $H_n$ a period of 17.2±0.1 Oe was found. This is in a good agreement with value ΔH=17.1 Oe calculated from the period of this array, which was determined using SEM. Detailed analysis of the data presented in Fig.1 already provides us with several important conclusions. As can be seen from the Fig.1b the dip for n=0 is very similar to the other dips. According to the results of Ref.16 this excludes a scenario involving interstitial vortices. For such a case n=0 corresponds to pinning on holes and higher numbers of n are related to interstitial sites with pinning induced by inter-vortex repulsion. Clearly, from the data presented in Fig.1b, one would expect the same mechanism of matching anomalies for all n. From the dip structure two characteristic fields may be distinguished. The resistivity starts to decrease in the field range $H_n \pm H^*$ with $H^* = 5 \div 7$ Oe . The dips are extremely sharp but using fine magnetic field steps of $5 \cdot 10^{-3}$ Oe, we find inflection points and a clear saturation in a region of $H_n \pm H_{Ln}$ as can be seen from the Fig.1c. The parameter $H_{Ln}$ ($H_{L0} \approx 0.2$ Oe for 7.76K) will be defined more accurately below. The dashed line shows the $\Delta R \propto (H - H_L)^{1/2}$ dependence[24] expected when approaching the locked state in the absence of thermal fluctuations. To elucidate this narrow field feature in more details we have analyzed the resistive response to ac modulations of the magnetic field.

The Fig.2a demonstrates the behavior of the first harmonic $U_{1H}$ of the ac signal as function of the magnetic field. The inset shows the region near the matching field on an expanded scale. The $U_{1H}$ value deviates from zero in the magnetic field interval $H_n \pm H^*$. Its most prominent feature is a very fast change in the sign of the response in an extremely narrow interval $H_n \pm H_{Ln}$ near the matching field. Taking into account the slope of the response curve of typically 100 μV/Oe and an intrinsic noise of 1nV/Hz$^{1/2}$ we can estimate the sensitivity to magnetic field as $10^{-9}$ T/Hz$^{1/2}$. We believe this could be improved further by optimization of array parameters.

To investigate the locked states further and find the values of $H_{Ln}$ more accurately we have analyzed the second harmonic $U_{2H}$ of the ac response to modulation of the magnetic field. As demonstrated in Fig.2b a series of very sharp peaks exist near the matching fields. The shape of these peaks is clearly shown on an extended scale in the insert of Fig.2b. For most field values the second harmonic is equal to zero or slightly negative (for $|H - H_n| < H^*$). However, close to the matching field large second and higher order harmonics appear in a narrow field range, which we relate to strong non-linearities induced by the locked state. Using zero points of $U_{2H}$ we can accurately estimate the width of the locked regions, for instance $H_{L1} \approx 0.25$Oe and less than 0.2Oe for n=0, which corresponds to 1% of the matching period.

To demonstrate these locked states by an independent technique we considered also the non-linearities in current-voltage characteristics. To detect such non-linearities we use the second harmonic of the response to an ac-current in a slowly varying magnetic field. As can be seen from Fig.3 far from the matching field the non-linearity is very small (the E-J curve is essentially linear) and near the matching field two different regimes of non-linearity can be detected. For the field interval $H_n \pm H^*$ a signal relatively slow varying with H is observed. In contrast to this, for the second type of non-linearity the second harmonic increases sharply in a narrow field interval. The arrows indicate the points where the second harmonic of the resistive response $U_{2H}$ (see Fig.2b) equals zero. We can see that the range of strong non-linearities in the E-J characteristics corresponds perfectly to the region of strong non-linearities in the R(H) dependences.

The results presented above clearly demonstrate the existence of very narrow locked states. Using the zero points in the second harmonic of the resistive response in modulation measurements, the width $H_L$ of these states was estimated for different commensurability numbers and temperatures. Fig.4a shows that the width of the locked state increases linearly with n. The temperature dependence is non-monotonic (Fig.4b). The sharpest states are realized near T=7.76 K. The increase of the width at low temperatures can be related to the increase of the condensation energy, which results in a larger hole pinning. At high temperatures the increase is probably due to thermal smearing. A similar broadening was observed for a commensurate Xe lattice on a carbon substrate[5].

The commensurability of a periodic elastic media with a periodic pinning potential has been studied since the 1930s. In the first approach, Frenkel and Kontorova[25] considered a one-dimensional chain. Later the 2D problem was addressed in several publications[26]. The vortex system in a periodical potential can be mapped to a system of quantum bosons[17], for which various models have been developed already [27]. Free quantum mechanical particles are delocalized and described by a combination of Bloch waves spreading through the whole system. When the particle density is commensurate with the periodic pinning potential a gap in the energy develops and the system forms a Mott insulator. For incommensurate filling, a transition to a superfluid phase of bosons was found[27]. Sufficiently strong disorder in the pinning potential produces a third phase: an insulating Bose glass. For a system with disordered columnar defects only a state for n=1 was found. Later Mott phases for other commensurate numbers were shown for periodic pinning potentials[18]. The Mott phase is an analogue of the Meissner phase. Both the tilt modulus and compression modulus $C_{11}$ should be infinite in the Mott insulator. However, some softening may result because of entropy effects[28]. The thermal activation via the energy gap of a Mott insulator should give rise to finite elastic moduli and a non-zero value of the resistivity.

Our results demonstrate clearly the existence of at least two types of states strongly interacting with the periodical pinning array. One is related to the broad region around the matching field and the second is localized very close to it. Obviously, when an integer number of the vortex lattice periods equals the period of pinning potential a commensurate state is realized. When the magnetic field changes slightly from a commensurate value the number of vortices tends to remain unchanged inducing a compression of the vortex lattice. After reaching a critical deformation value the first soliton-like particle appears. Depending on the intrinsic parameters this might be either a change in the flux number of multiquanta vortex by one or an interstitial vortex. The locked phase is incompressible and represents the analogue of the Meissner phase in bulk superconductors. The self-energy of a single n-quanta vortex is $E_n = (n\Phi_o / 4\pi\lambda)^2 \ln(\lambda/R)$, where R is the radius of the hole[29]. The field for creation of a vortex excitation in the homogeneous multiquantum locked state may be estimated as $H_{c1n} = (E_{n+1} - E_n) / \Phi_o = (2n+1) H_{c10}$ similarly to the lower critical field $H_{c1}$ in bulk superconductors. This explains the linear dependence of the width of the locked state, which we observe in our experiments (Fig.4a). For large deviations from the matching field, $|H - H_n| > H^*$, the concentration of soliton excitations $((B - B_n) d^2 / \Phi_o)$ exceeds 30-40% of the value for holes in the superconducting film, which probably corresponds to a percolation threshold and full delocalization of solitons. Our interpretation in terms of multi-quanta vortices is in agreement with recent observations using scanning Hall probe imaging[30]. These results demonstrated very clearly a homogeneous locked state at the matching field. A shift by only 7.5% from the matching field induced a large inhomogeneity characteristic for a de-localized state. Our measurements provide evidence for the existence of a locked state in a narrow field interval, reveal the boundary $H_L$ for the lock-in transition and its variations with the temperature and matching order.

Summarizing, the electric losses in a periodic array of holes in a superconducting Nb film have been studied. Our results demonstrate the existence of extremely narrow states at the matching fields, which result in a non-linear modulation response and in a non-linear current-voltage characteristics. The width of these states changes with temperature and commensurability numbers. We relate these sharp anomalies to a locked commensurate state of the Mott insulator type.

**FIGURE CAPTIONS**

Fig.1. Field dependence of the resistance for three different temperatures T=7.72, 7.76 and 7.80 K. Fig.1b&c show dips for n=0, ±1 on an expanded scale for T=7.76 K. Arrows indicate characteristic fields H* and $H_{Ln}$ for the n=0 state. The dashed line shows the expected behavior[24] at the locked state in the absence of thermal fluctuations.

Fig.2. Field dependence of the first (a) and second (b) harmonics of the ac response in a field modulation experiment at T=7.76 K ($H_{ac}$=50 mOe, f=68 Hz, $I_{dc}$=100 µA). The inset shows the behavior near n=0 (a) and n=1 (b) on an expanded scale.

Fig.3. Field dependence of the second harmonic response in the sinusoidal current modulation experiment with amplitude 100 µA (68 Hz) at T=7.76 K. It characterizes the non-linearity of the current voltage characteristics. The inset shows the shape of the curve near n=1 on an expanded scale. Arrows indicate the region of the locked state found from $U_{2H}$ (Fig.2b).

Fig.4. Width of the locked states vs. commensurability number (a) and temperature (b). Lines are guides to the eye.

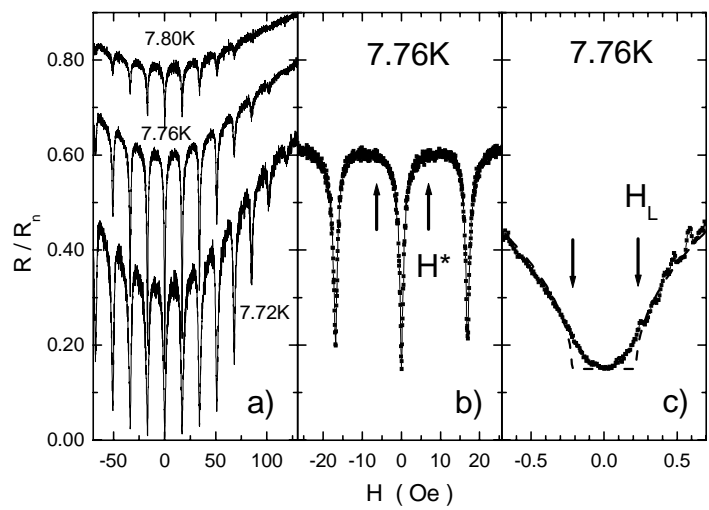

Fig.1

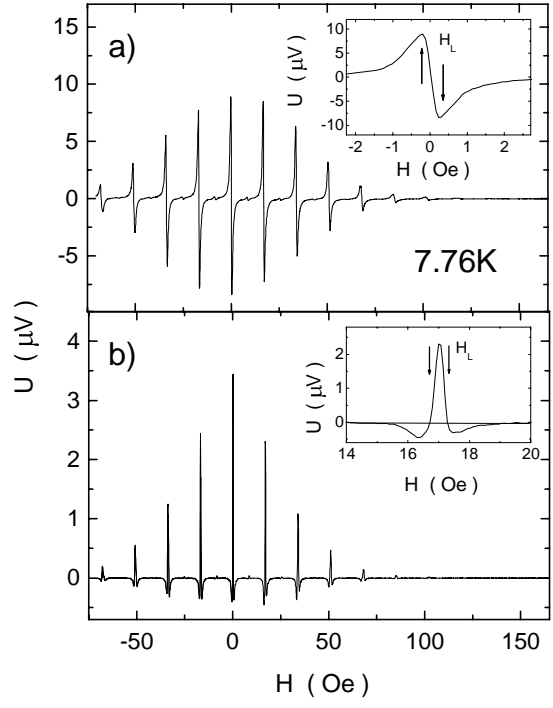

Fig.2

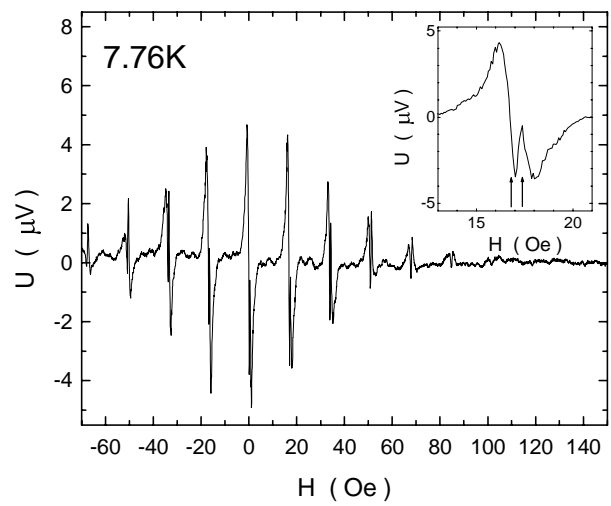

Fig.3

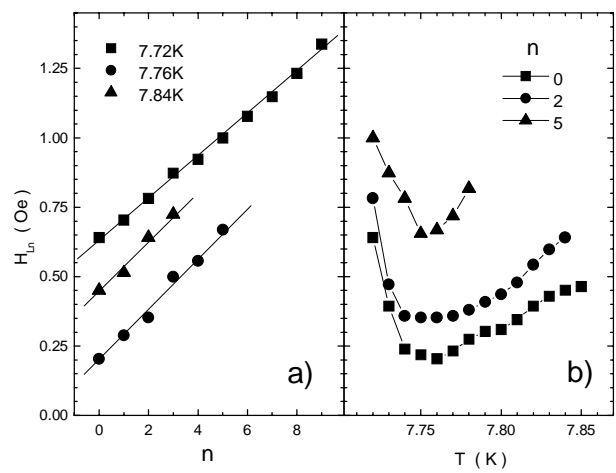

Fig.4